\documentclass{PoS}

\usepackage{cite}

\newcommand{\Br}{\mathrm{Br}}

\def\be{\begin{equation}}
\def\ee{\end{equation}}
\def\beqn{\begin{eqnarray}}
\def\eeqn{\end{eqnarray}}

\title{Effects of a charged Higgs boson in $B \rightarrow D^{(*)} \tau \nu$ decays}

\ShortTitle{Effects of a charged Higgs boson in $B \rightarrow D^{(*)} \tau \nu$ decays}

\author{\speaker{Alejandro Celis}%
        \thanks{I would like to thank Martin Jung, Xin-Qiang Li and Antonio Pich for the collaboration leading to the results presented in this talk.  The work of A. C. is supported by the Spanish Ministry MECD through the FPU grant AP2010-0308. }\\
        IFIC, Universitat de Val\`encia -- CSIC, Apt. Correus 22085, E-46071 Val\`encia, Spain\\
       E-mail: \email{alejandro.celis@ific.uv.es}}

\abstract{In this talk I consider the current status of $B \rightarrow D^{(*)} \tau \nu$ decays in the context of two-Higgs-doublet models.  The BaBar collaboration has reported an excess in semileptonic $b \rightarrow c \tau \nu$ transitions that has gathered recent interest as a possible signature of new physics.   The sensitivity of these decays to tree-level charged Higgs boson contributions is discussed,  taking into account the constraints from other semileptonic and leptonic meson decays in which the charged Higgs would enter at the same level.     }

\FullConference{The European Physical Society Conference on High Energy Physics \\
		 18-24 July, 2013\\
		 Stockholm, Sweden}

\begin{document}

\section{Introduction}

The universality of the leptonic charged-current couplings has been tested experimentally at the $0.2\%$ level~\cite{Beringer:1900zz}.  A recent measurement of the exclusive semileptonic $\bar B\rightarrow D^{(*)} \tau^- \bar \nu_{\tau}$ decay rates performed by the BaBar collaboration, suggesting a significant violation of lepton flavour universality (LFU), has raised interest as a possible new physics signature.   The BaBar collaboration measured the ratios~\cite{Lees:2012xj}
\begin{equation} \label{eq1}
R(D^{(*)}) =  \frac{\Gamma( \bar B \rightarrow D^{(*)} \tau^- \bar \nu_{\tau}  )}{\Gamma( \bar B \rightarrow D^{(*)} \ell^- \bar \nu_{\ell}  ) } \,,
\end{equation}
 where $\ell$ is either $e$ or $\mu$, finding $R(D) = 0.440 \pm 0.058 \pm 0.042$ and $R(D^*) = 0.332 \pm 0.024 \pm 0.018$.  These results show a $2.0 \sigma$ and  $2.7 \sigma$ excess with respect to the standard model (SM) predictions respectively~\cite{Kamenik:2008tj,Fajfer:2012vx}.  Subsequent theoretical works have found a better agreement between the SM and the experimental value for $R(D)$ by revisiting the treatment of the hadronic matrix elements.  Unquenched lattice QCD calculations have obtained a SM prediction that is $\sim 1 \sigma$ higher than previous estimates for $R(D)$~\cite{Bailey:2012jg} and similarly a new calculation of $R(D)$ trying to minimize the theory input finds an agreement between the SM and the experiment within $2 \sigma$~\cite{Becirevic:2012jf}.    
 
 Though the excess in $R(D)$ observed by the BaBar collaboration basically disappears when considering the most recent theoretical estimates of this quantity~\cite{Bailey:2012jg,Becirevic:2012jf}, the situation for $R(D^*)$ is still unclear.    Certainly, an update by the Belle collaboration using the full dataset is needed to further asses the significance of this excess.   
 
 It is interesting to consider what kind of NP could explain the present experimental value for $R(D^{*})$ while being compatible with other flavour constraints.   I will focus on a minimal extension of the SM scalar sector by the addition of a second Higgs doublet, a two-Higgs-doublet model (2HDM).   The scalar spectrum contains in this case a charged scalar that could give rise naturally to the observed violation of LFU in these transitions.   For a recent review of the two-Higgs-doublet model literature, see~Ref.~\cite{Branco:2011iw}.   
    In this talk I will show that to accommodate the latest BaBar measurement of $R(D^*)$ within this model, a departure from the family universality of the Yukawa couplings is required.  With this purpose in mind, the current excess in $R(D^{(*)})$ is analyzed within the Aligned two-Higgs-doublet model (A2HDM)~\cite{Pich:2009sp}, taking into account other meson decays in which the charged Higgs enters also at tree level.   The processes considered are listed in Table \ref{tab::SM}, a more detailed discussion of the results presented here can be found in Refs.~\cite{Jung:2010ik,Celis:2012dk}.

\section{Charged Higgs effects in $B \rightarrow D^{(*)} \tau \nu$ decays}
Charged Higgs interactions with fermions  in the A2HDM are described by~\cite{Pich:2009sp}
\beqn\label{lagrangianHp}
 \mathcal L  \supset &-& \frac{\sqrt{2}}{v}\; H^+ \left\{ \bar{u} \left[ \varsigma_d\, V M_d \mathcal P_R - \varsigma_u\, M_u^\dagger V \mathcal P_L \right]  d\, + \, \varsigma_l\, \bar{\nu} M_l \mathcal P_R l \right\}
\;  + \;\mathrm{h.c.} \,.
\eeqn
Here the family-universal alignment parameters $\varsigma_{f= u,d,l}$ are in general complex and independent, thus introducing new sources of CP-violation beyond the SM.  Note also that the only source of flavour-changing phenomena is the CKM matrix $V$.   In Eq.~(\ref{lagrangianHp}) we have: $v = ( \sqrt{2} G_F)^{-1/2} \simeq 246$~GeV, $M_{f =u,d,l}$ denote the diagonal mass matrices of the fermions and $\mathcal P_{L,R} = (1 \mp  \gamma_5)/2$ are the usual chirality projectors.  The more restricted versions of the 2HDM  in which a $Z_2$ symmetry is imposed in order to eliminate tree-level flavour-changing neutral currents (FCNCs), can be recovered as particular limits of this parametrization, see Table~\ref{tab:models}.  
\begin{table}[h]\begin{center}
\begin{tabular}{|c|c|c|c|}
\hline
Model & $\varsigma_d$ & $\varsigma_u$ & $\varsigma_l$  \\
\hline
Type I  &  $\cot{\beta}$ & $\cot{\beta}$ &  $\cot{\beta}$ \\
Type II &  $-\tan{\beta}$ &  $\cot{\beta}$ &  $-\tan{\beta}$ \\
Type X  &  $\cot{\beta}$ &  $\cot{\beta}$ &  $-\tan{\beta}$ \\
Type Y  &  $-\tan{\beta}$ &  $\cot{\beta}$ &  $\cot{\beta}$ \\
Inert  & 0 & 0 & 0 \\
\hline
\end{tabular}
\caption{2HDMs with natural flavour conservation. }
\label{tab:models}
\end{center}\end{table}
The latest BaBar measurements of $R(D)$ and $R(D^*)$ exclude the type-II 2HDM at $99.8\%$~CL for any value of $\tan \beta/M_{H^{\pm}}$ as long as $M_{H^{\pm}} > 10$~GeV~\cite{Lees:2012xj}.     The remaining region of the parameter space, namely $M_{H^{\pm}} \leq 10$~GeV, is excluded in Ref.~\cite{Lees:2012xj} by considering the loop-induced process $\bar B \rightarrow X_s \gamma$, which sets the $\tan \beta$-independent bound $M_{H^{\pm}} > 380$~GeV at $95\%$~CL in this model~\cite{Hermann:2012fc}.   While this argument is perfectly correct, it relies on a comparison of limits obtained from tree-level and loop-induced processes.     Loop-induced processes like $\bar B \rightarrow X_s \gamma$ are much more sensitive than those mediated at tree-level to a possible UV completion of the theory, heavy degrees of freedom can enter in the loops at the same level than the charged Higgs causing interfering contributions.   One would also like to consider more general Yukawa structures within the 2HDM framework beyond the type-II, for which the $\bar B \rightarrow X_s \gamma$ process is not as restrictive.  In the type-I 2HDM for example, a $95\%$ CL bound on the charged Higgs mass which is independent of $\tan \beta$ cannot be extracted from $\bar B \rightarrow X_s \gamma$~\cite{Hermann:2012fc}.

Constraints on a possible charged scalar from LEP offer an alternative way to complement the limits from $R(D^{(*)})$ without relying on loop-induced processes.   A charged Higgs with mass below $80$~GeV has been excluded in the type-II scenario at $95\%$~CL for any value of $\tan \beta$ by direct searches performed at LEP~\cite{Abbiendi:2013hk}, under the assumption that the charged Higgs only decays into fermions.   A more general limit can also be considered; for very light charged Higgs bosons, pair-production of charged scalars in $e^+ e^-$ collisions via the s-channel exchange of a $Z$ boson would modify the $Z$ decay width.  Given that the $Z H^+ H^-$ vertex is completely fixed by the gauge symmetry, precision measurements of the $Z$-width at LEP then translate into an indirect bound on the charged Higgs mass, $M_{H^{\pm}} > 39.6$~GeV at $95\%$~CL, which does not depend on the charged Higgs decay channels or the Yukawa structure~\cite{Abbiendi:2013hk}.

\begin{figure}[ht]
\centering
\includegraphics[width=7.4cm,height=7.4cm]{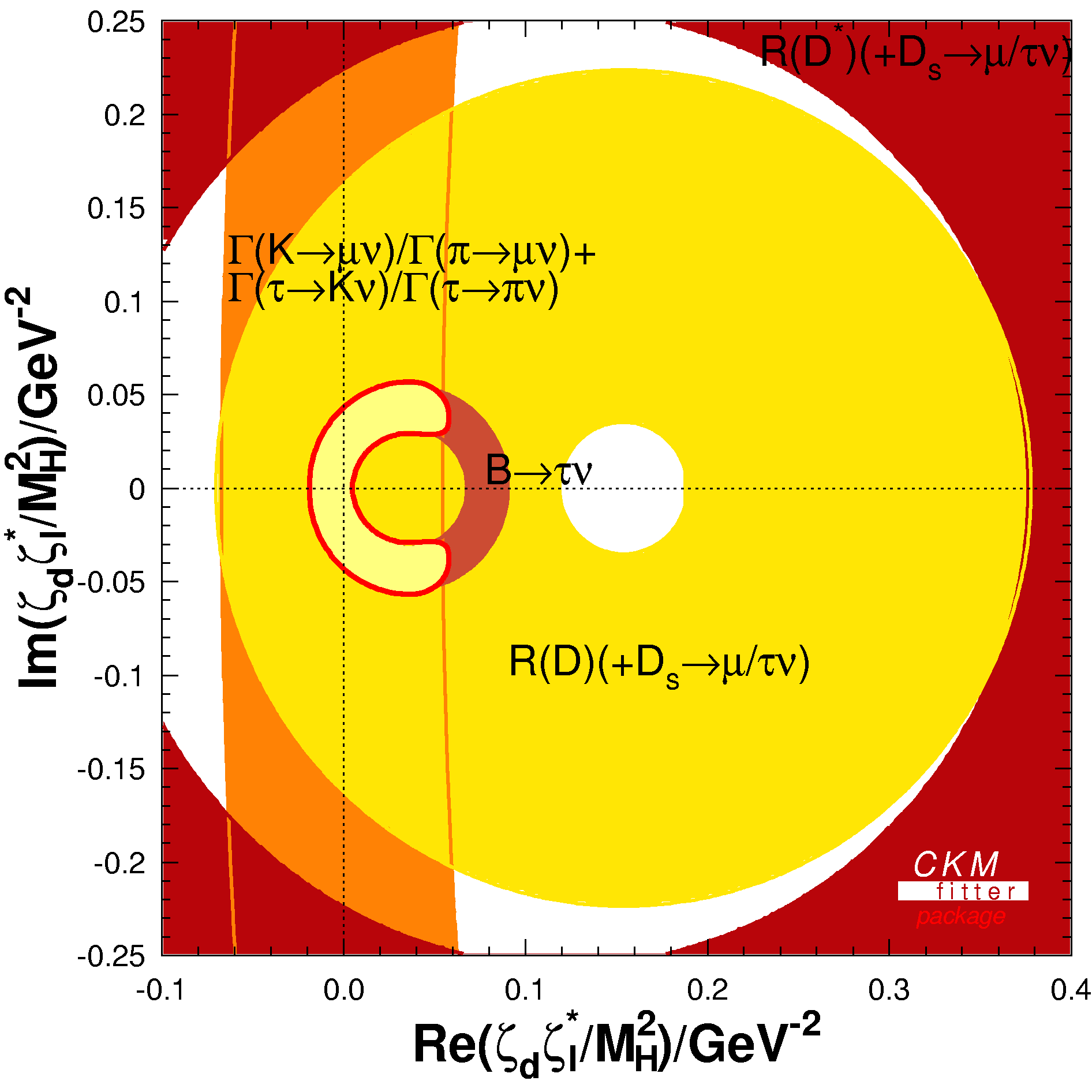}
~
\includegraphics[width=7.4cm,height=7.4cm]{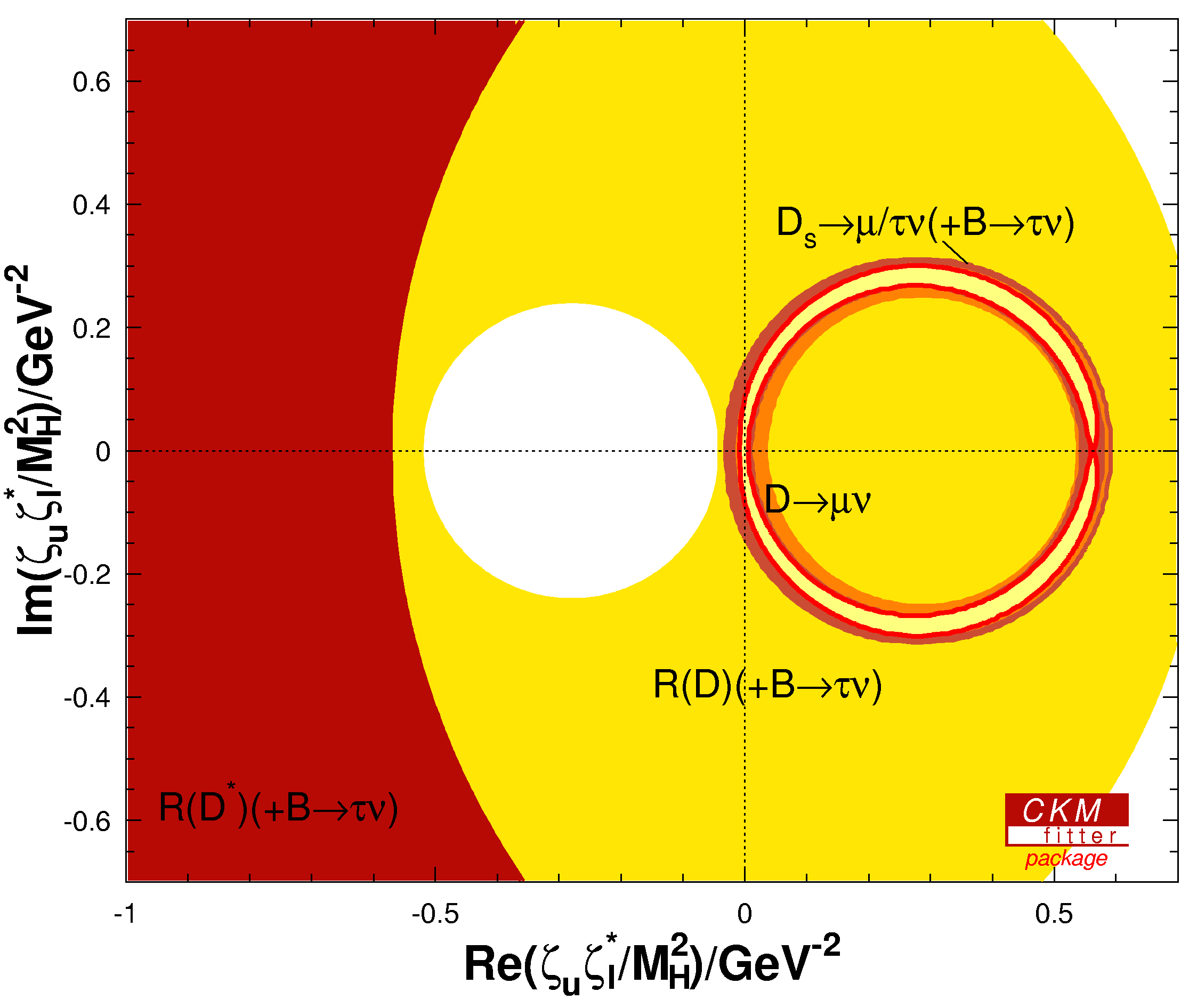}
\caption{\label{allowed} \it \small   Constraints in the complex planes $\varsigma_d \varsigma_l^*/M_{H^{\pm}}^2$ (left) and  $\varsigma_u \varsigma_l^*/M_{H^{\pm}}^2$ (right), in units of GeV$^{-2}$, from various leptonic and semileptonic decays.    Allowed regions at $95\%$~CL are shown for different combinations of observables.   }
\end{figure}

An interesting place where a charged Higgs can also induce violations of LFU is in leptonic $B \rightarrow \tau \nu$ decays.   In order to avoid the ambiguity in the determination of $|V_{ub}|$, it is convenient to consider the ratio~\cite{Fajfer:2012jt}
\be
R_{\tau/\ell}^{\pi} \equiv  \frac{\tau(B^0)}{\tau(B^-)} \frac{\mathrm{Br}(  B^- \rightarrow \tau^- \bar \nu)}{\mathrm{Br}(\bar B^0 \rightarrow \pi^+ \ell^- \bar \nu )}  = 0.73\pm 0.15 \,,
\ee
where the quoted experimental value considers the average $\Br(B\to \tau \nu_\tau) = (1.14 \pm 0.23)\times 10^{-4}$.  The SM prediction for this observable is $R_{\tau/\ell}^{\pi} = 0.31(6)$~\cite{Fajfer:2012jt}.  It was first found in Ref.~\cite{Fajfer:2012jt} that none of the 2HDMs with natural flavour conservation (NFC) can simultaneously accommodate the current measurements of the three ratios $\{R(D),R(D^{*}),R_{\tau/\ell}^{\pi}\}$.   To accommodate the experimental data on tauonic-$B$ decays: $B \rightarrow D^{(*)} \tau \nu$ and $B \rightarrow \tau \nu$, one has to allow for a more general Yukawa structure as provided by the Yukawa alignment hypothesis (A2HDM) or by assuming the presence of tree-level FCNCs~\cite{Crivellin:2012ye}.      The preferred values for the charged Higgs couplings needed to explain  $R(D^*)$ and $\mathrm{Br}(B \rightarrow \tau \nu)$ within the A2HDM, are however not compatible with current constraints from $D_{(s)}$-meson leptonic decay rates as shown in Fig.~\ref{allowed}~\cite{Celis:2012dk}. 

   In Fig.~\ref{allowed} (right-plot) it can be seen for example that $R(D^*) + B \rightarrow \tau \nu$ prefer large and negative values for $\mathrm{Re}(\varsigma_u \varsigma_l^*)/M_{H^{\pm}}^2$, while $D_{(s)}  + B \rightarrow \tau \nu$ prefers a positive $\mathrm{Re}(\varsigma_u \varsigma_l^*)/M_{H^{\pm}}^2$.  An important tension between these observables can also be observed in the $(\varsigma_d \varsigma_l^*)/M_{H^{\pm}}^2$ plane.  The experimental data and SM predictions of all the observables considered in this figure are given in Table~\ref{tab::SM}, note that all these processes have in common that receive tree-level charged scalar contributions.   In order to explain the present experimental value for $R(D^*)$ within 2HDMs it is then necessary to assume a departure from the family universality of the Yukawa couplings.

\begin{table}[ht]
\begin{center}
\doublerulesep 0.7pt \tabcolsep 0.07in
\small{
\begin{tabular}{lccc}
\hline\hline
Observable   				&  SM Prediction					&  Experimental Value   \\
\hline
 $R({D})$ 							& $0.296^{+0.008}_{-0.006} \pm 0.015$ & $ 0.438 \pm 0.056$ \\
 $R({D^*})$       					& $0.252 \pm 0.002\pm0.003$ 		& $0.354 \pm 0.026$		 \\ 
 $\Br(B\to \tau \nu_\tau)$ 	&  $(0.79^{+0.06}_{-0.04} \pm 0.08)\times 10^{-4}$& $(1.15 \pm 0.23)\times 10^{-4}$	 \\
 $\Br(D_s \to \tau \nu_\tau)$	&  $(5.18 \pm 0.08\pm0.17) \times 10^{-2} $& $(5.54 \pm 0.24)\times 10^{-2}$ \\
 $\Br(D_s \to \mu \nu)$	& $(5.31 \pm 0.09\pm0.17) \times 10^{-3} $	& $(5.54 \pm 0.24)\times 10^{-3}$	 \\
 $\Br(D \to \mu \nu)$	&  $(4.11^{+0.06}_{-0.05}\pm0.27) \times 10^{-4} $ & $(3.76 \pm 0.18)\times 10^{-4}$	 \\
 $\Gamma(K\to\mu\nu)/\Gamma(\pi\to\mu\nu)$    & $1.333\pm0.004\pm0.026$ & $1.337\pm0.003$  \\
 $\Gamma(\tau\to K\nu_\tau)/\Gamma(\tau\to\pi\nu_\tau)$ & $(6.56\pm0.02\pm0.15)\times10^{-2}$ & $(6.46\pm0.10)\times10^{-2}$ \\
 \hline\hline
 \end{tabular}}
\caption{\label{tab::SM} \it \small  SM predictions and experimental values for the different observables considered in the analysis, see Ref.~\cite{Celis:2012dk} for details.  The first uncertainty corresponds to the statistical error, and the second, when present, to the theoretical error.}
\end{center}
\end{table}

Several observables can be constructed in $B \rightarrow D^{(*)} \tau \nu $ decays taking advantage of the rich three-body kinematics and the spin structure of the final state (due to the $\tau$-lepton and the $D^*$), for a recent discussion see Refs.~\cite{Fajfer:2012vx,Celis:2012dk,Datta:2012qk,Tanaka:2012nw,Biancofiore:2013ki,Duraisamy:2013pia}.  These observables can be useful to disentangle different kinds of new physics in $b \rightarrow c \tau \nu$ transitions and more importantly to our discussion, they can be used to test for charged scalar contributions independently of the flavour structure of its fermionic couplings.    One obvious combination of observables that is not sensitive to a charged scalar is~\cite{Celis:2012dk}
\begin{equation}\label{eq::Xa}
X_1(q^2) \equiv  R_{D^*}(q^2)-R_L^*(q^2) \,,
\end{equation}
given that the charged Higgs does not contribute to the transverse helicity amplitudes. Here $q^{2}=(p_{B}-p_{D^{(*)}})^2$ is the momentum transfer squared and $R_L^*(q^2)$ denotes the differential longitudinal decay rate normalized by the light lepton mode
\begin{equation} \label{eq:RLproc}
R_L^*(q^2) =  \frac{d \Gamma^{L}_{\tau}/d q^2}{d \Gamma^{L}_{\ell}/d q^2} \,.
\end{equation}
Another observable of this kind can be constructed~\cite{Celis:2012dk}
\begin{equation}\label{eq::X1b}
X_2^D(q^2) \equiv  R_D(q^2)\,(A^{D}_{\lambda}(q^2)+1)\quad\mbox{and}\quad X_2^{D^*}(q^2) \equiv  R_{D^{*}}(q^2)\,(A^{D^{*}}_{\lambda}(q^2)+1) \,,
\end{equation}
using the $\tau$ spin asymmetry defined in the $\tau$-$\bar\nu_{\tau}$ center-of-mass frame  
\begin{equation} \label{eq:ATAUDproc}
A^{D^{(*)}}_{\lambda}(q^2) = \frac{ d\Gamma^{D^{(*)}}[\lambda_{\tau} =  -1/2]/ d q^2  - d\Gamma^{D^{(*)}}[\lambda_{\tau} = + 1/2] /d q^2 }{ d\Gamma^{D^{(*)}}[\lambda_{\tau} =  -1/2]/ d q^2  + d\Gamma^{D^{(*)}}[\lambda_{\tau} = + 1/2] /d q^2}\,,
\end{equation}
where $\lambda_{\tau}=\pm 1/2$ denotes the helicity of the final $\tau$-lepton.   In case any deviation from the SM prediction is observed in the future for $X_1(q^2)$ or $X_2^{D^{(*)}}(q^2)$, it would be a clear indication of non-scalar new physics.   Note also that the previous observables are clean from charged scalar contributions in the hole $q^2$ range and can be subject to a $q^2$-binned analysis.  CP-violating observables with these properties can also be defined~\cite{Duraisamy:2013pia}.

\section{Conclusions}
While recent theoretical predictions for $R(D)$ in the SM are in better agreement with the present experimental average~\cite{Bailey:2012jg,Becirevic:2012jf}, the situation for $R(D^*)$ still needs to be clarified.   A future update of this modes by the Belle collaboration using the full dataset is of utmost importance at this point.   I have shown in this talk how current measurements of $R(D^*)$, $\mathrm{Br}(B \rightarrow \tau \nu)$ and $D_{(s)}$-meson leptonic decay rates cannot be accommodated within two-Higgs-doublet models without tree-level flavour-changing neutral currents.  This conclusion of course also holds for all the different versions of the 2HDM with natural flavour conservation, being just particular cases of the parametrization considered here.    

The decays $B \rightarrow D^{(*)} \tau \nu$ will continue playing an important role in the long term.   So far we have only accessed a small amount of the information these processes can give us, namely the ratios $R(D^{(*)})$ and information about the $q^2$ spectra~\cite{Lees:2013uzd}.    Future Super-Flavour factories could in principle exploit further the rich kinematics and spin-structure in $ B\rightarrow D^{(*)} \tau  \nu$ decays, not available for two-body meson decay modes.

\end{document}